\documentclass[aps,pre,amsmath,amsfonts,amssymb,superscriptaddress,showpacs,11pt]{revtex4}
\usepackage{graphics}
\usepackage[latin1]{inputenc}
\usepackage{amscd}
\usepackage{soul}

\usepackage[pdftex]{graphicx}

\usepackage{rotating}
\usepackage[pdftex,hypertexnames=true,linktocpage=true]{hyperref}

\usepackage{color}

\usepackage{amsthm}
\usepackage{framed}
\usepackage{amssymb}
\usepackage{amsmath,mathrsfs}
\usepackage{hhline}
\usepackage{ulem}
\usepackage{fancyhdr}
\usepackage{tikz}
\everymath{\displaystyle}
\everymath{\displaystyle}
\everymath{\displaystyle}
\newcommand{\beq}{\begin{equation}}
\newcommand{\eeq}{\end{equation}}

\newcommand{\bi}{\begin{itemize}}
\newcommand{\ei}{\end{itemize}}

\newcommand{\Hilb}{\mathcal{H}}

\def\RR{\mathbb{R}}

\def\CC{\mathbb{C}}

\newcommand{\Tau}{\mathcal{T}}

\newcommand{\Ma}{\mathrm{M}}
\newcommand{\esse}{\mathcal{S}}

\newcommand{\Div}{\mathcal{D}}
\newcommand{\nihat}{\mathrm{X}}
\newcommand{\grad}{\mathrm{grad}}

\newcommand{\Riemann}{\mathcal{R}}
\newcommand{\tangent}{{\rm T}}

\newcommand{\paralleltransport}{{\rm P}}

\newcommand{\metric}{{\rm g}}

\newcommand{\bxi}{\boldsymbol{ \xi}}

\newcommand{\Tr}{{\rm Tr}}

\begin{document}

\title{\textbf{Canonical divergence for flat $\alpha$-connections: Classical and Quantum}}
\author{Domenico Felice}
\email{felice@mis.mpg.de}
\affiliation{Max Planck Institute for Mathematics in the Sciences\\
 Inselstrasse 22--04103 Leipzig,
 Germany}
  \author{Nihat Ay}
\email{nay@mis.mpg.de}
\affiliation{ Max Planck Institute for Mathematics in the Sciences\\
 Inselstrasse 22--04103 Leipzig,
 Germany\\
Santa Fe Institute, 1399 Hyde Park Rd, Santa Fe, NM 87501, USA\\
 Faculty of Mathematics and Computer Science, University of Leipzig, PF 100920, 04009 Leipzig, Germany}

\begin{abstract}
{A recent canonical divergence, which is introduced on a smooth manifold $\Ma$ endowed with a general dualistic structure $(\metric,\nabla,\nabla^*)$, is considered for flat $\alpha$-connections.} In the classical setting, we compute such a canonical divergence on the manifold of positive  measures and prove that it coincides with the classical $\alpha$-divergence. In the quantum framework, the recent canonical divergence is evaluated for the quantum $\alpha$-connections on the {manifold of all positive definite  Hermitian operators.} Also in this case we obtain that the recent canonical divergence is the quantum $\alpha$-divergence.
\end{abstract}


\maketitle 
 
\section{Introduction}

Methods of Information Geometry (IG) \cite{Amari00} are ubiquitous in physical sciences and encompass both classical and quantum systems \cite{Felice18}.  
The natural object of study in IG is a quadruple $(\Ma,\metric,\nabla,\nabla^*)$ given by a smooth manifold $\Ma$, a Riemannian metric  $\metric$ and a pair of affine connections on $\Ma$ which are dual with respect to $\metric$,
\begin{equation}
\label{dualconnections}
X\,\metric\left(Y,Z\right)\,=\,\metric\left(\nabla_X Y,Z\right)+\metric\left(Y,\nabla_X^*Z\right)\,,
\end{equation}
for all sections $X,Y,Z\in\Tau(\Ma)$. The quadruple $(\Ma,\metric,\nabla,\nabla^*)$ is called {\it statistical manifold} whenever the dual connections are both torsion-free \cite{Ay17}. Actually, the notion of statistical manifold, introduced by Lauritzen \cite{Lauritzen87}, is usually referred to the triple $(\Ma,\metric,\mathrm{T})$, where $\mathrm{T}(X,Y,Z)=\metric\left(\nabla_X^*Y-\nabla_X Y,Z\right)$ is a $3$-symmetric tensor. However, when $\nabla$ and $\nabla^*$ are both torsion-free connections, the structures $(\Ma,\metric,\nabla,\nabla^*)$ and $(\Ma,\metric,\mathrm{T})$ are equivalent \cite{Ay17}. When $\Ma$ is a manifold of probability distributions, $\metric\equiv\metric^{\mathrm{F}}$ is the Fisher metric, and $\nabla\equiv\nabla^{(e)}$ and $\nabla^*\equiv\nabla^{(m)}$ are the exponential and mixture connections \cite{Amari82}, IG has been
successfully applied to many fields, such as statistical inference, control systems theory, and neural networks (see \cite{Amari16} and references quoted therein for a comprehensive literature on  applications of IG).

The geometric structure of a statistical manifold is encoded by a smooth function $\Div:\Ma\times\Ma\rightarrow\RR$ such that
\begin{equation}
\label{Divergence}
\Div(p,q)\geq 0\,,\qquad \mbox{and}\quad \Div(p,q)=0\quad\mbox{iff}\quad p=q\,,
\end{equation}
for all $p,q\in\Ma$ \cite{Eguchi85}. The dualistic structure $(\metric,\nabla,\nabla^*)$ of $\Ma$ is then recovered in the following way:
\begin{eqnarray}
\label{metricfromdiv}
&&\metric_{ij}(p)=-\left.\partial_i\partial_j^{\prime} \Div(\bxi_p,\bxi_q)\right|_{p=q}=\left.\partial^{\prime}_i\partial_j^{\prime} \Div(\bxi_p,\bxi_q)\right|_{p=q}\\
\label{connectiondfromdiv}
&& \Gamma_{ijk}(p)=-\left.\partial_i\partial_j\partial_k^{\prime} \Div(\bxi_p,\bxi_q)\right|_{p=q}, \qquad {\Gamma}^*_{ijk}(p)=-\left.\partial^{\prime}_i\partial^{\prime}_j\partial_k \Div(\bxi_p,\bxi_q)\right|_{p=q}\,.
\end{eqnarray} 
Here, 
\begin{equation*}
\partial_i=\frac{\partial}{\partial \xi_p^i} \quad \mbox{and}\quad \partial^{\prime}_i=\frac{\partial}{\partial \xi_q^i}
\end{equation*} 
and $\{\bxi_p:=(\xi_p^1,\ldots,\xi_p^n)\}$ and $\{\bxi_q:=(\xi_q^1,\ldots,\xi_q^n)\}$ are local coordinate systems of $p$ and $q$, respectively. Moreover, $\Gamma_{ijk}=\metric\left(\nabla_{\partial_i}\partial_j,\partial_k\right)$, ${\Gamma}^*_{ijk}=\metric\left(\nabla^*_{\partial_i}\partial_j,\partial_k\right)$ are the symbols of the dual connections $\nabla$ and $\nabla^*$, respectively. The function $\Div$ is called a {\it divergence or contrast function} of the statistical manifold $(\Ma,\metric,\nabla,\nabla^*)$ \cite{Eguchi92}.

The function $\Div$ is called a {\it flat divergence} if the dualistic structure $(\metric,\nabla,\nabla^*)$ introduced on a smooth manifold $\Ma$ by Eqs. (\ref{metricfromdiv}) and (\ref{connectiondfromdiv}) is flat, namely the curvature tensors $\Riemann(\nabla)$ and $\Riemann^*(\nabla^*)$ are zero. In this particular case, an attempt to connect IG with physics has been established in \cite{Fujiwara95}. There, the connection between IG and integrable dynamical systems has been bridged by  a divergence, which is {\it canonical} in some sense. More precisely, for $q\in\Ma$ the gradient flows $\dot{\bxi}=-\grad_{\bxi}\,\Div(\bxi_q,\cdot)$ and $\dot{\bxi}=-\grad_{\bxi}\,\Div(\cdot,\bxi_q)$ converge to the point $q$ along the $\nabla^*$-geodesic and the $\nabla$-geodesic, respectively. In this context, when $\Ma$ is the manifold of Gaussian distributions with mean $\mu$ and variance $\sigma^2$, the Kullback-Leibler (KL) divergence induces a dualistic structure given by the Fisher metric and the $(e),\,(m)$ connections. In this case, the dynamics of the above mentioned gradient flows, given now in terms of the KL divergence, turns out to be the Uhlenbeck-Ornstein process $\mu=\mu_0+v(t+\tau)\,,\,\, \sigma^2=2\,D(t+\tau)$ characterized by the drift coefficient $v$ and diffusion coefficient $D$ \cite{Fujiwara95}. Further connections between IG and dynamical systems can be found in \cite{Nakamura93} where certain gradient flows on Gaussian and multinomial distributions
are characterized as completely integrable Hamiltonian systems.

The Kullback-Leibler divergence has been effectively employed to quantify the complexity of a system described by a probability distribution $p$ in terms of its deviation from an exponential family of probability distributions \cite{Aycomplexity}. The quantum version of the KL divergence, namely the {\it quantum relative entropy}, induces on the manifold of quantum states a dually flat structure given by a quantum version of the Fisher metric tensor and two flat connections, also called the mixture and the exponential ones, which are dual in the sense of Eq. (\ref{dualconnections}) \cite{Nagaoka95}. Moreover, the quantum relative entropy has been used to quantify the many-party correlations of a composite quantum state $\rho$ as the deviation of it from a Gibbs family of quantum states \cite{Ayquantum}. The effectiveness of the quantum relative entropy as measure of complexity for quantum states has been showed in \cite{Niekamp}, where algorithms for its evaluation are studied. Furthermore, in that context, the  many-party correlations is related to the entanglement of quantum systems as defined in \cite{Vedral}.

A generalization of the flat structures induced by the Kullback-Leibler divergence on the finite classical systems and by the quantum relative entropy on the finite quantum states is provided by the classical $\alpha$-divergence and the quantum $\alpha$-divergence, respectively. Both generate a $1$-parameter family of connections, the {\it $\alpha$-connections}, which are dual with respect to the Fisher metric in the classical case \cite{Amari00} whereas in the quantum case they are dual with respect to the Wigner-Yanase-Dyson metric \cite{Grasselli04}. 

From a physical viewpoint, it is worth to remark that the Boltzmann-Gibbs distribution in statistical physics is an exponential family such that an invariant flat structure is given to the underlying manifold in terms of the $(e)$-connection \cite{Amari16}. Tsallis generalized the concept of Boltzmann-Gibbs distribution by introducing a generalized entropy, called the {\it $\mathrm{q}$-entropy}, for studying various phenomena not included in the conventional Boltzmann-Gibbs framework \cite{Tsallis,Tsallis19}. Actually, the $\alpha$-geometry, which is induced by the $\alpha$-divergence, covers the geometry of $\mathrm{q}$-entropy physics \cite{Ohara}. Therefore, the $\alpha$-divergence can be understood as a generalization of the classical KL divergence also from a physical standpoint. In the quantum case, the geometry of $\mathrm{q}$-entropy physics has been successfully employed for carrying out a criterium that detects the critical frontier which has separable states on one side and quantum entangled ones on the other one \cite{Abe}.

The purpose of the present article is to consider a recent canonical divergence, introduced by Ay and Amari in \cite{Ay15}, as tool for unifying classical and quantum Information Geometry. In particular, we aim to prove that this canonical divergence is the classical $\alpha$-divergence when computed on the space of  positive measures as well as the quantum $\alpha$-divergence if evaluated on the manifold of positive definite  Hermitian operators.

\section{Canonical Divergence and inverse problem in Information Geometry}\label{CanDic}

The inverse problem within Information Geometry  concerns the search for a divergence function $\Div$ which recovers a given dual structure $(\metric,\nabla,\nabla^*)$ of a smooth manifold $\Ma$ according to Eqs. (\ref{metricfromdiv}) and (\ref{connectiondfromdiv}). {The solution to this probelm was provided by  {Matumoto} who  showed that such a divergence always exists for any statistical manifold} \cite{matumoto1993}. {Nonetheless, this solution is not unique and  infinitely many divergences can be defined on $\Ma$ which give the same dualistic structure. For this reason, seeking out a divergence  that  can be  considered as {\it the most natural} is of utmost importance. To this end, Amari and Nagaoka defined a Bregman type divergence on dually flat manifolds} \cite{Amari00}. {This one has  relevant properties such as the generalized Pythagorean theorem and the geodesic projection theorem and it is
 commonly assessed as the natural solution of the inverse problem in Information Geometry for dually flat manifolds. This is exaclty why the Amari and Nagaoka divergence is referred to as {\it the canonical divergence} of dually flat statistical manfiolds.} {However, the need for a general canonical divergence, which applies to any dualistic structure, is a very crucial issue, as pointed out in \cite{AyTusch}. {According to the theory developed in \cite{Ay17}, a divergence function $\Div$ of a statistical manifold $(\Ma,\metric,\nabla,\nabla^*)$ is called {\it canonical} if:
\begin{itemize}
\item $\Div$ generates the dualistic structure $(\metric,\nabla,\nabla^*)$ based on Eqs. \eqref{metricfromdiv}, \eqref{connectiondfromdiv};
\item $\Div$ is one half of the squared Riemannian distance, i.e. $\Div(p,q)=\frac{1}{2}\,d(p,q)^2$, when the statistical manifold is self-dual, namely when $\nabla=\nabla^*$ coincides with the Levi--Civita connection of $\metric$;
\item $\Div$ is the canonical divergence of Bregman type when $(\Ma,\metric,\nabla,\nabla^*)$ is dually flat.
\end{itemize}  
}

{Ay and Amari have  recently defined a divergence for a general dualistic structure in terms of geodesic integration of the inverse exponential map \cite{Ay15}. It turns out that such a divergence satisfies all the requirements above mentioned. Therefore, it can be viewed as a canonical divergence for a general dualistc structure. For $p,q\in\Ma$ consider the $\nabla$-geodesic $\widetilde{\gamma}(t)\,(0\leq t\leq 1)$ connecting $q$ with $p$, the recent canonical divergence introduced in \cite{Ay15} is then defined by}
\begin{equation}
\label{AyDiv}
\Div(p,q):=\int_0^1\left\langle\nihat_t(p),\dot{\widetilde{\gamma}}(t)\right\rangle_{\widetilde{\gamma}(t)}\,\mathrm{d} t\,,\quad \nihat_t(p):=\exp_{\widetilde{\gamma}(t)}^{-1}(p)\,.
\end{equation}
{The $\nabla$-exponential map, $\exp:\tangent\Ma\rightarrow\Ma$,  is defined by $\exp(X)=\gamma_X(1)$ whenever the $\nabla$-geodesic $\gamma_X(t)$, {satisfying $\dot{\gamma}_X(0)=X$,} exists on an interval of $t$ containing $[0,1]$.  Therefore, if $\gamma:[0,1]\rightarrow\Ma$ is the $\nabla$-geodesic such that $\gamma(0)=p$ and $\gamma(1)=q$, the inverse at $p$ of the exponential map is given by $\exp_p^{-1}(q):=\dot{\gamma}(0)$. According to this definition, for every $t\in [0,1]$ we can consider the $\nabla$-geodesic $\gamma_t(s)$ such that $\gamma_t(0)=p$ and $\gamma_t(1)=\gamma(t)$ and then define the $\nabla$-velocity vector at $p$ by $\nihat_p(\gamma(t)):=\exp_p^{-1}(\gamma(t))=\dot{\gamma}_t(0)$. In this way, the vector field $\nihat_t(p)$ of the Eq. \eqref{AyDiv} turns out to be given by $\nihat_t(p)=\paralleltransport_{\gamma(t)}\,\nihat_p(\gamma(t))=t\,\dot{\gamma}(t)$. Here, $\paralleltransport:\tangent_p\Ma\rightarrow\tangent_{\gamma(t)}\Ma$ is the $\nabla$-parallel transport from $p$ to $\gamma(t)$.} In the light of all this, the divergence $\Div(p,q)$ assumes the following useful expression:
\begin{equation}
\label{AyDivnice}
\Div(p,q)=\int_0^1\,t\,\|\dot{\gamma}(t)\|^2\,d t\,.
\end{equation}

Analogously, the dual function of $\Div(p,q)$ is defined as the $\nabla^*$-geodesic integration of the inverse of the $\nabla^*$-exponential map \cite{Ay15}. Therefore, we have for the dual divergence $\Div^*$ a similar expression as the Eq. (\ref{AyDivnice}) for the canonical divergence $\Div$:
\begin{equation}
\label{AyDivnice*}
\Div^*(p,q)=\int_0^1\,t\,\|\dot{\gamma}^*(t)\|^2\,d t\,,
\end{equation}
where $\gamma^*(t)\,(0\leq t\leq 1)$ is the $\nabla^*$-geodesic connecting $p$ with $q$.

The canonical divergence given by Eq. (\ref{AyDivnice}) has been recently proposed as tool for unifying classical and quantum Information Geometry \cite{Felice19}. In particular, it has been considered on the simplex of probability measures
\begin{equation}
\label{simplex}
\esse:=\left\{p=\sum_i\,p_i\,\delta_i\in\RR^{n}\,|\,p_i>0\,\mbox{for all}	\,i,\,\mbox{and}\,\sum_i\,p_i=1\right\}\,,
\end{equation}
as well as on the manifold of quantum finite states,
\begin{equation}
\label{quantumstates}
\mathrm{S}:=\left\{\rho:\mathcal{H}\rightarrow\CC\,|\,\rho=\rho^{\dagger}>0,\,\Tr\rho=1\right\}\,,
\end{equation}
where $\mathcal{H}$ denotes a finite-dimensional Hilbert space and $\rho$ is any Hermitian operator on $\mathcal{H}$. The natural dualistic structure on the simplex $\esse$ is given, in classical Information Geometry, in terms of the Fisher metric $\metric^{\mathrm{F}}$ and two flat connections, the mixture $\nabla^{(m)}$ and the exponential $\nabla^{(e)}$ ones, which are dual with respect to $\metric^{\mathrm{F}}$ in the sense of Eq. (\ref{dualconnections}) \cite{Amari82}. Very remarkably, {the Fisher metric is the only {\it monotone}  Riemannian metric (up to a positive factor) on the class of finite probability simplices \cite{Centsov}. The quantum version of a monotone Riemannian metric on the manifold of quantum finite states $\mathrm{S}$ is given in terms of the notion of stochastic mapping \cite{Petz&Sudar}. More precisley, let $\mathcal{A}$ the set of all Hermitian operators on the Hilbert space $\Hilb$, a linear mapping $T:\mathcal{A}\rightarrow\mathcal{A}$ is said to be {\it stochastic} if $T(\mathrm{S})\subset\mathrm{S}$ and $T$ is completely positive \cite{Holevo}. Furthermore, a family $\{\langle\langle\, ,\,\rangle\rangle_{\rho}\,|\,\rho\in\mathrm{S}\}$ of inner products on $\mathcal{A}$ is said to be monotone if $\langle\langle A,A\rangle\rangle_{\rho}\geq\langle\langle T(A),T(A)\rangle\rangle_{T(\rho)}$ for any arbitrary stochastic map $T$ and for every $\rho\in\mathrm{S}$ \cite{Petz&Sudar}. Due to Petz there are infinitely many monotone inner products on $\Hilb$. Therefore, the quantum analogue of the Fisher metric is not unique  \cite{Petz96}.}  However, when the flat connections are required torsion-free, a natural dualistic structure on the manifold of quantum states $\mathrm{S}$ is the one induced by the Bogoluibov-Kubo-Mori (BKM) inner product \cite{Nagaoka95}. Furthermore, it turns out that the only monotone metrics which make the mixture connection $\nabla^{(m)}$ and the exponential connection $\nabla^{(e)}$ dual are the scalar multiples of the BKM metric \cite{Grasselli01}. When the canonical divergence (\ref{AyDivnice}) is computed on the simplex $\esse$, it is showed to be the Kullback-Leibler divergence,
\begin{equation}
\label{KL}
\Div(p,q)\,=\, \sum_{i=1}^{n}\,p_i\,\log\left(\frac{p_i}{q_i}\right)\,,\quad p,\,q\,\in\esse\,,
\end{equation}
which proves that $\Div(p,q)$ recovers the natural dualistic structure on $\esse$ given by the Fisher metric $\metric^{\mathrm{F}}$ and the dually flat connections $\nabla^{(m)}$ and $\nabla^{(e)}$ \cite{Felice19}. Analogously, when $\Div$ is considered on the manifold of quantum states $\mathrm{S}$, it has been proved that 
\begin{equation}
\label{quantumrelativentropy}
\Div(\rho,\sigma)=\Tr\,\rho\left(\log\rho-\log\sigma\right)\,,\quad \rho,\,\sigma\,\in\mathrm{S}\,,
\end{equation}
where $\Tr$ denotes the trace operator on the finite-dimensional Hilbert space of density matrices \cite{Felice19}. The function on the right-hand-side of Eq. (\ref{quantumrelativentropy}) is called the {\it quantum relative entropy} and it recovers the dual structure of $\mathrm{S}$ given by the metric induced by the BKM inner product and the flat connections $\nabla^{(m)}$ and $\nabla^{(e)}$ \cite{Amari00}.

{In this article, we aim to investigate} the canonical divergence (\ref{AyDivnice}) on the manifold of positive measures as well as on the space of positive definite Hermitian operators for the more general $\alpha$-connections. In the classical Information Geometry, the one parameter family of the $\alpha$-connections is defined on the manifold of positive measures by the linear combination of the mixture and exponential connections \cite{Amari16},
\begin{equation}
\label{alphaclassical}
\nabla^{\alpha}\,=\,\frac{1-\alpha}{2}\, \nabla^{(m)}+\frac{1+\alpha}{2}\,\nabla^{(e)}\,.
\end{equation}
It turns out that $\nabla^{\alpha}$ and $\nabla^{-\alpha}$ are dual with respect to the Fisher metric $\metric^{\mathrm{F}}$ in the sense of Eq. (\ref{dualconnections}). In quantum Information Geometry, $\alpha$-connections appeared also in terms of the Amari $\alpha$-embeddings \cite{Hasegawa97}. {While in the classical Information Geometry the two definitions coincide, this is no longer true in quantum Information Geometry \cite{Jencova01}. In the present paper, we consider the definition of the quantum $\alpha$-connections by means of the Amari $\alpha$-embedding which excludes that they can be obtained by the convex mixture of $\nabla^{(m)}$ and $\nabla^{(e)}$ connections. The natural inner product that makes the quantum $\alpha$-connections dual in the sense of Eq. \eqref{dualconnections} is the Wigner-Yanase-Dyson (WYD) metric which appeared for the first time in the context of quantum Information Geometry in the work of Hasegawa \cite{Hasegawa92}. Actually, it turns out that this is the only monotone metric (up to a scalar multiuple) that makes the quantum $\nabla^{\alpha}$ and $\nabla^{-\alpha}$ dual \cite{Grasselli04}.}

\section{Classical flat alpha-divergence}
We represent measures on the set $\{1,\ldots,n\}$ as elements of $\RR^n$. In this representation, the Dirac measures $\delta_i,\,i=1,\ldots,n$, form the canonical basis of $\RR^n$. The $n$-dimensional cone of positive measures on the set  $\{1,\ldots,n\}$ is then defined by
\begin{equation}
\label{cone}
\mathcal{M}_+:=\RR^n_{+}=\left\{p=\sum_{i=1}^n\,p_i\,\delta_i\in\RR^n\,|\, p_i>0\,\,\mbox{for all}\,\, i\right\}\,.
\end{equation}
The very natural Riemannian metric on $\mathcal{M}_+$ is the Fisher metric \cite{Amari16} which is defined by
\begin{equation}
\label{Fishermetric}
\metric_p^{\mathrm{F}}(X,Y)=\sum_{i=1}^n\, \frac{1}{p_i}\,X_i\,Y_i\,,
\end{equation}
for all $p\in\mathcal{M}_+$ and $X,Y\in\tangent_p\mathcal{M}_+$. Here, $\tangent_p\mathcal{M}_+$ denotes the tangent space to $\mathcal{M}_+$ at $p$. Given the mixture and the exponential connections we can define the $\alpha$-connection on $\mathcal{M}_+$ by using Eq. (\ref{alphaclassical}). Recalling that $\tangent_p\mathcal{M}_+\equiv\RR^n$, we can write any $X\in\tangent_p\mathcal{M}_+$ with respect to the canonical basis of $\RR^n$. Hence, the $(m)$-connection on $\mathcal{M}_+$ reads as follows
\begin{equation}
\label{mconnection}
\left.\nabla^{(m)}_X Y\right|_p=\sum_{i=1}^n\frac{\partial Y_i}{\partial X}(p)\,\delta_i\,,
\end{equation}
for all $X=(X_1,\ldots,X_n)$ and $Y=(Y_1,\ldots,Y_n)$ in the tangent space $\tangent_p\mathcal{M}_+$. Here, $\partial/\partial X$ denotes the derivative in the direction $X\in\tangent_p\mathcal{M}_+$. On the contrary, the $(e)$-connection is given by
\begin{equation}
\label{econnection}
\left.\nabla^{(e)}_X Y\right|_p=\sum_{i=1}^n\left(\frac{\partial Y_i}{\partial X}(p)-\frac{1}{p_i}\,X_i\,Y_i\right)\,\delta_i\,.
\end{equation}
Therefore, by applying Eq. (\ref{alphaclassical}) we can describe the $\alpha$-connection for the manifold $\mathcal{M}_+$ as follows,
\begin{eqnarray}
\label{alphaconneconnectionclassical}
\left.\nabla_X^{\alpha}Y\right|_p &=&\frac{1-\alpha}{2} \left.\nabla^{(m)}_X Y\right|_p+\frac{1+\alpha}{2}\left.\nabla^{(e)}_X Y\right|_p\nonumber\\
&=& \left.\nabla^{(m)}_X Y\right|_p+\frac{1+\alpha}{2}\left(\left.\nabla^{(e)}_X Y\right|_p-\left.\nabla^{(m)}_X Y\right|_p\right)\nonumber\\
&=& \sum_{i=1}^n\left(\frac{\partial Y_i}{\partial X}(p)-\frac{1+\alpha}{2}\frac{X_i\,Y_i}{p_i}\right)\,\delta_i\,,
\end{eqnarray}
for all $X,\,Y\in\tangent_p\mathcal{M}_+$. It turns out that the dualistic structure $(\metric^{\mathrm{F}},\nabla^{\alpha},\nabla^{-\alpha})$ is dually flat, i.e. the Riemannian curvature tensors of $\nabla^{\alpha}$ and $\nabla^{-\alpha}$ are zero \cite{Amari16}. Furthermore, very naturally, such an $\alpha$-flat dualistic structure is induced on $\mathcal{M}_+$ by the  $\alpha$-divergence $D^{(\alpha)}$ which is commonly assessed as the canonical solution of the inverse problem for recovering the dually flat $\alpha$-structure $(\metric^{\mathrm{F}},\nabla^{\alpha},\nabla^{-\alpha})$ on $\mathcal{M}_+$ \cite{Amari16}. Given $p,\,q\in\mathcal{M}_+$, the $\alpha$-divergence between $p$ and $q$ is given by
\begin{equation}
\label{alphadivergenceC}
D^{(\alpha)}(p,q)\,=\,\sum_{i=1}^n\,\left(\frac{2}{1-\alpha}\,q_i+\frac{2}{1+\alpha}\,p_i-\frac{4}{1-\alpha^2}\,q_i^{\frac{1+\alpha}{2}}p_i^{\frac{1-\alpha}{2}}\right)\,.
\end{equation}
{The flat $\alpha$-divergence \eqref{alphadivergenceC} has been deeply studied and its features have been widely discussed in literature (see \cite{Ay17}, \cite{Amari16} for more details). In particular, we point out that  $D^{(\alpha)}$ is a continuous function of the parameter $\alpha$ and the limit $\alpha\rightarrow -1$ gives the well-known  
 Kullback-Leibler divergence on the manifold of positive measures,
\begin{equation}
\label{alphaKL}
\lim_{\alpha\rightarrow-1} D^{(\alpha)}(p,q)=\sum_{i=1}^n \left(q_i-p_i-p_i\log\frac{q_i}{p_i}\right)\,,\quad p,q\in\mathcal{M}_+\,,
\end{equation}
whereas the  limit $\alpha\rightarrow+1$ gives 
$$
\lim_{\alpha\rightarrow+1} D^{(\alpha)}(p,q)=\sum_{i=1}^n \left(p_i-q_i-q_i\log\frac{p_i}{q_i}\right)\,,\quad p,q\in\mathcal{M}_+\,.
$$
Let us observe that, if we restrict \eqref{alphaKL} to the simplex of probability distributions, namely when $\sum_{i=1}^n\,p_i=\sum_{i=1}^n\,q_i=1$, then we obtain the function \eqref{KL}.
At this point, it is worth mentioning the close connection between the $\alpha$-divergence \eqref{alphadivergenceC} and the Tsallis relative entropy, or {\it $\mathrm{q}$-divergence}, on the manifold of positive measures. The $\mathrm{q}$-divergence on $\mathcal{M}_+$ is defined by (see \cite{Martins09} for more details)
\begin{equation}
\label{TsallisDiv}
D_{\mathrm{q}}(p,q)=\frac{1}{1-\mathrm{q}}\sum_{i=1}^n\left(\mathrm{q}\,p_i+(1-\mathrm{q})\,q_i-p_i^{\mathrm{q}}\,q_i^{1-\mathrm{q}}\right)\,,\quad p,q\in\mathcal{M}_+\,.
\end{equation}
By setting $\alpha=1-2\mathrm{q}$ we can easily verify that the flat $\alpha$-divergence \eqref{alphadivergenceC} and the Tsallis $\mathrm{q}$-divergence \eqref{TsallisDiv} coincide up to a scaling factor. Further, the  limit $\mathrm{q}\rightarrow 1$, that is $\alpha\rightarrow-1$, recovers the Kullback-Leibler divergence,
$$
\lim_{\mathrm{q}\rightarrow 1}\, D_{\mathrm{q}}(p,q)=\lim_{\alpha\rightarrow-1}\,\frac{1-\alpha}{2}\,D^{(\alpha)}(p,q)=\sum_{i=1}^n \left(q_i-p_i-p_i\log\frac{q_i}{p_i}\right)\,,\quad p,q\in\mathcal{M}_+\,.
$$
}

\vspace{.2cm}

In this section we aim to compute the canonical divergence (\ref{AyDivnice}) on $\mathcal{M}_+$ for the $\alpha$-connections given by (\ref{alphaconneconnectionclassical}). In order to achieve this result, we need to consider the geodesic with respect to the $\nabla^{\alpha}$-connection. Let $p,\,q\in\mathcal{M}_+$, a curve $\gamma:[0,1]\rightarrow\mathcal{M}_+$ from $p$ to $q$  is an $\alpha$-geodesic iff $\nabla^{\alpha}_{\dot{\gamma}}\dot{\gamma}=0$ and $\gamma(0)=p$, $\gamma(1)=q$. From Eq. (\ref{alphaconneconnectionclassical}) we then obtain the following geodesic equations
\begin{equation}
\label{geodeqs}
\ddot{\gamma}_i-\frac{1+\alpha}{2}\,\frac{\dot{\gamma}_i^2}{\gamma_i}=0,\,i=1,\ldots,n\,,\quad \gamma(0)=p,\,\,\gamma(1)=q\,,
\end{equation}
where $\gamma(t)=(\gamma_1(t),\ldots,\gamma_n(t))$. Hence, the solution of (\ref{geodeqs}) is the $\alpha$-geodesic from $p$ to $q$ and it is given by
\begin{equation}
\label{alphageodesic}
\gamma^{(\alpha)}(t)=\left((1-t)\,p^{\frac{1-\alpha}{2}}+t\,q^{\frac{1-\alpha}{2}}\right)^{\frac{2}{1-\alpha}},\,\,\, t\in[0,1]\,.
\end{equation}
At this point we can apply Eq. (\ref{AyDivnice}) to compute the canonical divergence $\Div(p,q)$ on the manifold $\mathcal{M}_+$ of  positive measures. From Eq. (\ref{Fishermetric}) we obtain
\begin{eqnarray*}
%\label{conenorm}
\|\dot{\gamma}^{(\alpha)}(t)\|^2_{\gamma^{(\alpha)}(t)}&=& \metric^{\mathrm{F}}_{\gamma^{(\alpha)}(t)}\left(\dot{\gamma}^{(\alpha)}(t),\dot{\gamma}^{(\alpha)}(t)\right)=\sum_{i=1}^n\,\frac{\left(\dot{\gamma}^{(\alpha)}_i\right)^2}{\gamma_i^{(\alpha)}}\\
&=&\sum_{i=1}^n\,\frac{4}{(1-\alpha)^2}\,\frac{\left((1-t)\,p^{\frac{1-\alpha}{2}}+t\,q^{\frac{1-\alpha}{2}}\right)^{2\frac{1+\alpha}{1-\alpha}}}{\left((1-t)\,p^{\frac{1-\alpha}{2}}+t\,q^{\frac{1-\alpha}{2}}\right)^{\frac{2}{1-\alpha}}}\,\left(q^{\frac{1-\alpha}{2}}-p^{\frac{1-\alpha}{2}}\right)^2\\
&=&\sum_{i=1}^n\,\frac{4}{(1-\alpha)^2}\,\left((1-t)\,p^{\frac{1-\alpha}{2}}+t\,q^{\frac{1-\alpha}{2}}\right)^{\frac{2\,\alpha}{1-\alpha}}\,\left(q^{\frac{1-\alpha}{2}}-p^{\frac{1-\alpha}{2}}\right)^2\,.
\end{eqnarray*}
Now, we can compute the integral in (\ref{AyDivnice}) by performing an integration by part:
\begin{eqnarray*}
%\label{integralcone}
\Div(p,q)&=&\sum_{i=1}^n\,\int_0^1\,t\,\frac{4}{(1-\alpha)^2}\left[\left((1-t)\,p_i^{\frac{1-\alpha}{2}}+t\,q_i^{\frac{1-\alpha}{2}}\right)^{\frac{2\,\alpha}{1-\alpha}}\,\left(q_i^{\frac{1-\alpha}{2}}-p_i^{\frac{1-\alpha}{2}}\right)\right]\,\left(q_i^{\frac{1-\alpha}{2}}-p_i^{\frac{1-\alpha}{2}}\right)\,\mathrm{d} t\\
&=&\sum_{i=1}^n\Bigg\{\left[t\frac{4}{(1-\alpha)^2}\frac{1-\alpha}{1+\alpha}\left((1-t)\,p_i^{\frac{1-\alpha}{2}}+t\,q_i^{\frac{1-\alpha}{2}}\right)^{\frac{1+\alpha}{1-\alpha}}\right]_0^1\left(q_i^{\frac{1-\alpha}{2}}-p_i^{\frac{1-\alpha}{2}}\right)\\
&&-\int_0^1\,\frac{4}{(1-\alpha)^2}\frac{1-\alpha}{1+\alpha}\left((1-t)\,p_i^{\frac{1-\alpha}{2}}+t\,q_i^{\frac{1-\alpha}{2}}\right)^{\frac{1+\alpha}{1-\alpha}}\left(q_i^{\frac{1-\alpha}{2}}-p_i^{\frac{1-\alpha}{2}}\right)\mathrm{d} t\Bigg\}\\
&=& \sum_{i=1}^n\frac{4}{1-\alpha^2}\left\{q_i^{\frac{1+\alpha}{2}}\left(q_i^{\frac{1-\alpha}{2}}-p_i^{\frac{1-\alpha}{2}}\right)-\int_0^1\left((1-t)\,p_i^{\frac{1-\alpha}{2}}+t\,q_i^{\frac{1-\alpha}{2}}\right)^{\frac{1+\alpha}{1-\alpha}}\left(q_i^{\frac{1-\alpha}{2}}-p^{\frac{1-\alpha}{2}}\right)\mathrm{d} t\right\}\\
&=& \sum_{i=1}^n\frac{4}{1-\alpha^2} \left(q_i-p_i^{\frac{1-\alpha}{2}}\,q_i^{\frac{1+\alpha}{2}}\right)-\frac{4}{1-\alpha^2}\frac{1-\alpha}{2}\left[\left((1-t)\,p_i^{\frac{1-\alpha}{2}}+t\,q_i^{\frac{1-\alpha}{2}}\right)^{\frac{2}{1-\alpha}}\right]_0^1\\
&=&\sum_{i=1}^n\left\{\frac{4}{1-\alpha^2}\,q_i-\frac{4}{1-\alpha^2}\,p_i^{\frac{1-\alpha}{2}}\,q_i^{\frac{1+\alpha}{2}}-\frac{2}{1+\alpha}\,q_i+\frac{2}{1+\alpha}\,p_i\right\}\,.
\end{eqnarray*}
This proves that
\begin{equation}
\Div(p,q)\,=\, \sum_{i=1}^n\,\left(\frac{2}{1-\alpha}\,q_i+\frac{2}{1+\alpha}\,p_i-\frac{4}{1-\alpha^2}\,q_i^{\frac{1+\alpha}{2}}p_i^{\frac{1-\alpha}{2}}\right)\,,
\end{equation}
that is the canonical divergence (\ref{AyDivnice}) corresponds to the $\alpha$-divergence (\ref{alphadivergenceC}) on the manifold $\mathcal{M}_+$ of  positive measures.

\section{Quantum flat alpha-divergence} 

The goal of this section is to compute the canonical divergence of Eq. \eqref{AyDivnice} on the manifold of  positive definite  Hermitian operators endowed with the quantum $\alpha$-connections. In the classical case, {the definition of the $\alpha$-connection $\nabla^{\alpha}$ can be equivalently given by Eq. \eqref{alphaclassical} or by means of the well-known Amari $\alpha$-embedding \cite{Amari16}. In the quantum setting, by exploiting the linear structure of the manifold we can introduce the quantum mixture connection $\nabla^{(m)}$ on the space of positive definite Hermitian operators. On the other hand, the quantum exponential connection $\nabla^{(e)}$ is introduced by relying on the linear structure of logarithms of the manifold. Both the connections, $\nabla^{(m)}$ and $\nabla^{(e)}$, are flat ones. Furthermore, they turn out to be dual in the sense of Eq. \eqref{dualconnections} with respect to the metric induced by the BKM inner product \cite{Nagaoka95}. Actually, this metric is (up to a scalar multiple) the only one that makes $\nabla^{(m)}$ and $\nabla^{(e)}$ dual on the manifold of positive definite Hermitian operators. A generalization of these two natural connections is provided by the quantum $\alpha$-connections which, on the  manifold under consideration, are flat ones, as well. In our approach, these $\alpha$-connections are introduced in terms of the Amari $\alpha$-embeddings and then they cannot be obtained by the convex combination of the mixture and the exponential connections as in the classical setting \cite{Grasselli04}. It turns out that for $\alpha\in[-3,3]$ the connections $\nabla^{\alpha}$ and $\nabla^{-\alpha}$ are torsion-free and dual with respect to the WYD metric \cite{Jencova03}. However, the target spaces of our approach are $L^r$, with $r=\frac{2}{1-\alpha}$, and then we restrict our discussion to the range $\alpha\in(-1,1)$. 

Recall that in order to compute the canonical divergence \eqref{AyDivnice}, we need to get the $\alpha$-geodesic $\gamma_{\alpha}$ between any two positive Hermitian operators $\rho_1$ and $\rho_2$. To do this, we shall describe the $\alpha$-connections in terms of the $\nabla^{\alpha}$-parallel transport which is introduced very naturally on the manifold of positive Hermitian operators.} Given $\mathcal{B}(\mathcal{H})$ the algebra of linear operators on  an $N$-dimensional complex Hilbert space a$\mathcal{H}$, the subspace of Hermitian operators is an $N^2$-dimensional real vector space defined by
\begin{equation}
\label{self-adjoint}
\mathcal{A}:=\{\rho\in\mathcal{B}(\mathcal{H})\,|\,\rho=\rho^{\dagger}\}\,,
\end{equation}
where $\rho^{\dagger}=\bar{\rho}^t$ and $t$, here, denotes the transpose matrix. Therefore, the manifold of all positive definite Hermitian operators or, more simply {\it quantum operators}, is given by
\begin{equation}
\label{Cone}
\mathrm{M}_{+}:=\{\rho\in\mathcal{A}\,|\,\rho>0\}\,.
\end{equation}
{The $\alpha$-embedding 
\begin{equation}
\label{alpha-embedding}
l_{\alpha}:\mathrm{M}_{+}\rightarrow\mathcal{A},\quad l_{\alpha}(\rho):=\frac{2}{1-\alpha}\,\rho^{\frac{1-\alpha}{2}}\,,
\end{equation}
maps the manifold of quantum operators into the vector space $\mathcal{A}$ for all $\alpha\in (-1,1)$ \cite{Amari00}. Furthermore, for $\rho\in\mathrm{M}_{+}$ we have that $\tangent_{\rho}\mathrm{M}_{+}=\tangent_{\rho}\mathcal{A}=\mathcal{A}$. Hence, the $\alpha$-embedding supplies a useful representation of the tangent bundle of $\mathrm{M}_{+}$. In fact, by considering the subspace} of $\mathcal{A}$ given by
\begin{equation}
\label{alpha-space}
\mathcal{A}^{(\alpha)}:=\left\{A\in\mathcal{A}\,|\, \Tr\left(\rho^{\frac{1-\alpha}{2}} A\right)=0\right\}\,,\qquad \rho\in\mathrm{M}_+\,,
\end{equation}
we can then define the isomorphism
\begin{equation}
\label{alpha-iso}
\left(l_{\alpha}\right)_{{*}(\rho)}:\tangent_{\rho}\mathrm{M}_{+}\rightarrow\mathcal{A}^{(\alpha)},\quad \left(l_{\alpha}\right)_{{*}(\rho)}(X):=\left(l_{\alpha}\circ\gamma\right)^{\prime}(0)\,,
\end{equation}
where $\gamma:(-\varepsilon,\varepsilon)\rightarrow\mathrm{M}_{+}$ is a curve such that $\gamma^{\prime}(0)=X$. This isomorphism provides the $\alpha$-representation of the tangent space $\tangent_{\rho}\mathrm{M}_{+}$ \cite{Grasselli04}. In particular, if $\left\{\xi^1,\ldots,\xi^n\right\}$ is a coordinate system for $\mathrm{M}_{+}$, then the $\alpha$-representation of the basis $\left\{\frac{\partial}{\partial\xi^1},\ldots,\frac{\partial}{\partial\xi^n}\right\}$ of $\tangent_{\rho}\mathrm{M}_{+}$ is $\left\{\frac{\partial l_{\alpha}}{\partial\xi^1},\ldots,\frac{\partial l_{\alpha}}{\partial\xi^n}\right\}$, where $n=N^2$.
Finally, for any vector field $X\in\Tau(\mathrm{M}_+)$ we have that its $\alpha$-representation is defined by 
\begin{equation}
\label{alpha-rep}
(X)^{(\alpha)}(\rho):=\left(l_{\alpha}\right)_{{*}(\rho)}X_{\rho}\,.
\end{equation}

\vspace{.3cm}

{From Eq. (\ref{alpha-embedding}) we may observe that the $l_{\alpha}(\rho)\in\mathrm{M}_+$ for all $\rho\in\mathrm{M}_+$. Now,  since $\tangent_{\rho}\mathrm{M}_{+}=\mathcal{A}$, we can simply define the $\alpha$-parallel transport on $\mathrm{M}_{+}$ by} 
\begin{equation}
\label{alpha-paralleltransport}
\Pi^{(\alpha)}_{\rho_1,\rho_2}:\tangent_{\rho_1}\mathrm{M}_{+}\rightarrow\tangent_{\rho_2}\mathrm{M}_{+},\quad \Pi^{(\alpha)}_{\rho_1,\rho_2}(X):=(l_{\alpha})_{*(\rho_2)}^{-1}\left((l_{\alpha})_{*(\rho_1)}(X)\right)\,,\quad \forall\,,\rho_1,\rho_2\in\mathrm{M}_+\,.
\end{equation}
Therefore, we find that the $\alpha$-representation of the covariant derivative $\nabla^{\alpha}$ associated to $\Pi^{(\alpha)}$ is
\begin{equation}
\label{alpha-covderiv}
\left(\nabla^{\alpha}_{\partial_i}\partial_j\right)^{(\alpha)}=\frac{\partial^2\,l_{\alpha}(\rho)}{\partial\xi^i\partial\xi^j}\,,
\end{equation}
where $\{\xi^1,\ldots,\xi^n\}$ is an arbitrary coordinate system for $\mathrm{M}_{+}$ and $n=N^2$ is the topological dimension of the tangent space $\tangent_{\rho}\mathrm{M}_+=\mathcal{A}$. In order to show that $\mathrm{M}_{+}$ is $\nabla^{\alpha}$-flat, consider a basis $\{A_1,\ldots,A_n\}$ of $\mathcal{A}$. Therefore, there exist real numbers $\{\theta^1,\ldots,\theta^n\}$ such that we can write  $\rho^{\frac{1-\alpha}{2}}\in\mathcal{A}$ as follows
$$
\frac{2}{1-\alpha}\rho^{\frac{1-\alpha}{2}}=\theta^1A_1+\ldots+\theta^n A_n\,,
$$
for every $\rho\in\mathrm{M}_+$. 
Then, we can see from Eq. (\ref{alpha-covderiv}) that
$$
\left(\nabla^{\alpha}_{\partial_i}\frac{\partial}{\partial\theta^j}\right)^{(\alpha)}=\frac{\partial^2 l_{\alpha}(\rho)}{\partial\theta^i\partial\theta^j}=\frac{\partial A_j}{\partial\theta^i}=0\,.
$$
This proves that $\{\theta^1,\ldots,\theta^n\}$ is a $\nabla^{\alpha}$-affine coordinate system for $\mathrm{M}_{+}$ and then $\mathrm{M}_{+}$ is $\nabla^{\alpha}$-flat \cite{Grasselli04}. As a consequence, for any couple of points $\rho_1,\rho_2\in\mathrm{M}_+$, we can write the $\alpha$-geodesic from $\rho_1$ to $\rho_2$  in this $\nabla^{\alpha}$-affine coordinates as follows
\begin{equation}
\label{alpha-geod}
\gamma_{\alpha}(t)=\frac{2}{1-\alpha}\left(t\,\rho_2^{\frac{1-\alpha}{2}}+(1-t)\,\rho_1^{\frac{1-\alpha}{2}}\right)\,.
\end{equation}

In order to perform the computation of the integral \eqref{AyDivnice}, we need to calculate the norm $\|\dot{\gamma}_{\alpha}(t)\|^2_{\gamma_{\alpha}(t)}$. To do this, we have to specify a suitable metric on the tangent space $\tangent\mathrm{M}_+$. As discussed above, we select the metric induced by the  WYD inner product because such a metric is the only one (up to a scalar multiple) that makes the quantum flat $\alpha$-connections dual in the sense of Eq. \eqref{dualconnections} \cite{Grasselli01}. For any $X,Y$ vector fields on $\mathrm{M}_+$, this metric turns out to be defined
by 
\begin{equation}
\label{WYDmetric}
\metric^{\alpha}_{\rho}\left(X,Y\right)=\Tr\left(X^{(\alpha)}\, Y^{(-\alpha)}\right),\quad X,Y\in\tangent_{\rho}\mathrm{M}_{+}\,,
\end{equation}
where $X^{(\alpha)}$ denotes the $\alpha$-representation of the tangent vector $X$ and it is given by Eq. \eqref{alpha-rep} \cite{Jancova02}. It is worth noticing that the  limit $\lim_{\alpha\rightarrow\pm 1} g^{\alpha}$ gives the metric induced by the BKM inner product on the manifold of density operators \cite{Nagaoka95}.

In order to write the WYD metric with respect to the
$\nabla^{\alpha}$-affine coordinate system $\{\theta^1,\ldots,\theta^n\}$, we may observe that
\begin{equation}
\label{alphaderiv}
\frac{\partial l_{\alpha}(\rho)}{\partial\theta^i}=A_i\,,
\end{equation}
where $A_i$ is a vector of the basis $\{A_1,\ldots,A_n\}\subset\mathcal{A}$. {This implies that, with this coordinate system, the $(\alpha)$-representation $X^{(\alpha)}$ of a vector field $X\in\Tau(\mathrm{M}_+)$ is the vector field $X$ itself.}
In addition, we also have that
\begin{equation}
\label{-alpha}
l_{-\alpha}=\frac{1+\alpha}{2}\rho^{\frac{1+\alpha}{2}}=\left(\frac{2}{1+\alpha}\right)\left(\frac{1-\alpha}{2}\right)^{\frac{1+\alpha}{1-\alpha}}l_{\alpha}^{\frac{1+\alpha}{1-\alpha}}\,.
\end{equation}
Therefore, in the $\alpha$-affine coordinate system $\{\theta^1,\ldots,\theta^n\}$, the components of the metric tensor \eqref{WYDmetric} are given by
\begin{eqnarray*}
\metric^{\alpha}_{ij}(\theta)&=&\Tr\left(\frac{\partial\,l_{\alpha}}{\partial \theta^i}\,\frac{\partial\,l_{-\alpha}}{\partial \theta^j}\right)\\
&=& \frac{2}{1-\alpha}\left(\frac{1-\alpha}{2}\right)^{\frac{1+\alpha}{1-\alpha}}\,\Tr\left(\frac{\partial\,l_{\alpha}}{\partial \theta^i}\,l_{\alpha}^{\frac{2\alpha}{1-\alpha}}\frac{\partial\,l_{\alpha}}{\partial\theta^j}\right)\,.
\end{eqnarray*}

The quantum dually flat structure $(\metric^{\alpha},\nabla^{\alpha},\nabla^{-\alpha})$ for the manifold of positive definite Hermitian operators (or quantum operators) so far described, can be also obtained through Eqs. \eqref{metricfromdiv} and \eqref{connectiondfromdiv} when the following divergence is considered, 
\begin{equation}
\label{QalphaDiv}
\mathrm{D}^{(\alpha)}(\rho_1,\rho_2):=\frac{4}{1-\alpha^2}\Tr\left(\frac{1-\alpha}{2}\,\rho_1+\frac{1+\alpha}{2}\,\rho_2-\rho_1^{\frac{1-\alpha}{2}}\rho_2^{\frac{1+\alpha}{2}}\right)\,, \quad \rho_1,\rho_2\in\mathrm{M}_+
\end{equation}
This function is called the {\it quantum $\alpha$-divergence} and {it has been introduced on $\mathrm{M}_+$ as the generalization of the $\alpha$-divergence \eqref{alphadivergenceC} of the positive measures \cite{Amari10}. Carrying on this line of reasoning, we can introduce a $\mathrm{q}$-divergence on the manifold of positive Hermitian operators, as well. In analogy with the classical case, we can set $\alpha=1-2\,\mathrm{q}$ in the argument of the trace $\Tr$ in the Eq. \eqref{QalphaDiv} and then we can write the following expression,
\begin{equation}
\label{Quantumqdivergence}
\mathrm{D}_{\mathrm{q}}(\rho_1,\rho_2):= \frac{1}{1-\mathrm{q}}\,\Tr\left[\mathrm{q}\,\rho_1+(1-\mathrm{q})\,\rho_2-\rho_1^{\mathrm{q}}\,\rho_2^{1-\mathrm{q}}\right]\,,\quad \rho_1,\rho_2\in\mathrm{M}_+\,,
\end{equation} 
which corresponds to the quantum $\alpha$-divergence up to the same scalar factor as in the classical case.
It is worth noting that this function is different from the extensions of the Tsallis relative entropy to the positive operators in the literature. For example, in \cite{Furuichi04} the quantum $\mathrm{q}$-divergence has been introduced in the following way,
\begin{equation}
\label{Tsallisquantumdivergence}
\widetilde{\mathrm{D}}_{\mathrm{q}}(\rho_1,\rho_2):=\frac{\Tr[\rho_1]-\Tr\left[\rho_1^{\mathrm{q}}\,\rho_2^{1-\mathrm{q}}\right]}{1-\mathrm{q}}\,,\quad \rho_1,\rho_2\in\mathrm{M}_+\,,
\end{equation}
for all $\mathrm{q}\in[0,1)$. However, both functions, $D_{\mathrm{q}}$ and $\widetilde{D}_{\mathrm{q}}$, reduces to the Tsallis relative entropy when restricted to the manifold $\mathrm{S}$ of density operators,
$$
\mathrm{D}_{\mathrm{q}}(\rho_1,\rho_2):=\frac{1-\Tr\left[\rho_1^{\mathrm{q}}\,\rho_2^{1-\mathrm{q}}\right]}{1-\mathrm{q}}\,,\quad \rho_1,\rho_2\in\mathrm{S}\,.
$$
Analogously, we can restrict the quantum $\alpha$-divergence \eqref{QalphaDiv} to the set of density operators and, since $\Tr\,\rho_1=\Tr\,\rho_2=1$, we can easily verify that \eqref{QalphaDiv} becomes
\begin{equation}
\label{alphaquantumdensity}
\mathrm{D}^{(\alpha)}(\rho_1,\rho_2)= \frac{4}{1-\alpha^2}\left(1-\Tr\left[\rho_1^{\frac{1-\alpha}{2}}\,\rho_2^{\frac{1+\alpha}{2}}\right]\right)\,,\quad \rho_1,\rho_2\in\mathrm{S}\,.
\end{equation}
Again, we can see that, on the manifold of density operators, the quantum $\alpha$-divergence coincides (up to a scalar factor) with the Tsallis relative entropy when we set $\alpha=1-2\,\mathrm{q}$. The quantum $\alpha$-divergence \eqref{alphaquantumdensity} was introduced and studied by Hasegawa in \cite{Hasegawa92} and it turns out to be a continuous function of the parameter $\alpha$. In particular, we have that
$$
\lim_{\alpha\rightarrow-1} \mathrm{D}^{(\alpha)}(\rho_1,\rho_2)=\Tr\left[\rho_1\left(\log\rho_1-\log\rho_2\right)\right]\,,
$$
that is the quantum relative entropy $\Div(\rho_1,\rho_2)$ as previously given in Eq. \eqref{quantumrelativentropy}. Furthermore, the  limit $\alpha\rightarrow+1$ gives $\lim_{\alpha\rightarrow+1}\mathrm{D}^{(\alpha)}(\rho_1,\rho_2)=\Div(\rho_2,\rho_1)$.
}

\vspace{.3cm}

At this point, we have in our hands all the ingredients necessary to compute the canonical divergence defined in \eqref{AyDivnice} between $\rho_1$ and $\rho_2$ on the manifold $\mathrm{M}_{+}$ of positive definite Hermitian operators,
\begin{equation}
\label{Aydivergence}
\Div(\rho_1,\rho_2):=\int_0^1\,t\,\|\dot{\gamma}_{\alpha}(t)\|^2_{\gamma_{\alpha}(t)}\,\mathrm{d} t\,,
\end{equation}
where $\gamma_{\alpha}(t)$ is the $\nabla^{\alpha}$-geodesic from $\rho_1$ to $\rho_2$. In the $\nabla^{\alpha}$-affine coordinates $\{\theta^1,\ldots,\theta^N\}$ this reads as in Eq. (\ref{alpha-geod}). Here, the norm $\|\cdot\|^2$ is induced by the WYD metric given in Eq. (\ref{WYDmetric}). Hence, in this case the canonical divergence \eqref{Aydivergence} can be written as
\begin{equation}
\label{alpha-div}
\Div(\rho_1,\rho_2)=\int_0^1 t\, \Tr\left((\dot{\gamma}_{\alpha}(t))^{(\alpha)}\,(\dot{\gamma}_{\alpha}(t))^{(-\alpha)}\right)\,\mathrm{d} t\,,
\end{equation}
where $(\dot{\gamma}_{\alpha}(t))^{(\alpha)}$ and $(\dot{\gamma}_{\alpha}(t))^{(-\alpha)}$ denote the $(\alpha)$ and the $(-\alpha)$ representations of $\dot{\gamma}(t)$. In order to compute the $\alpha$ representations of $\dot{\gamma}(t)$, we consider  Eq. (\ref{alpha-rep}) together with Eqs. (\ref{alphaderiv}), (\ref{-alpha}). Therefore, we have
\begin{eqnarray*}
(\dot{\gamma}_{\alpha}(t))^{(\alpha)} &=&\dot{\gamma}_{\alpha}(t)\\
 (\dot{\gamma}_{\alpha}(t))^{(-\alpha)}&=& \frac{2}{1-\alpha}\left(\frac{1-\alpha}{2}\right)^{\frac{1+\alpha}{1-\alpha}}\frac{1+\alpha}{1-\alpha}\,\left(\gamma_{\alpha}(t)\right)^{\frac{2\alpha}{1-\alpha}}\dot{\gamma}_{\alpha}(t)\\
 &=& \left(\frac{1-\alpha}{2}\right)^{\frac{2\alpha}{1-\alpha}}\,\left(\gamma_{\alpha}(t)\right)^{\frac{2\alpha}{1-\alpha}}\dot{\gamma}_{\alpha}(t)\,.
\end{eqnarray*}
We can plug these expressions in Eq. (\ref{alpha-div}). Hence, by performing an integration by parts, we get
\begin{eqnarray*}
\Div(\rho_1,\rho_2)&=&\left(\frac{1-\alpha}{2}\right)^{\frac{2\alpha}{1-\alpha}}\int_0^1 t\, \Tr\left(\dot{\gamma}_{\alpha}(t)\left(\gamma_{\alpha}(t)\right)^{\frac{2\alpha}{1-\alpha}}\dot{\gamma}_{\alpha}(t)\right)\,\mathrm{d} t\\
&=&\left(\frac{1-\alpha}{2}\right)^{\frac{2\alpha}{1-\alpha}}\Tr\left(\int_0^1 t\, \dot{\gamma}_{\alpha}(t)\left(\gamma_{\alpha}(t)\right)^{\frac{2\alpha}{1-\alpha}}\dot{\gamma}_{\alpha}(t)\,\mathrm{d} t\right)\\
&=&\left(\frac{1-\alpha}{2}\right)^{\frac{2\alpha}{1-\alpha}}\Tr\left(\left[t\dot{\gamma}_{\alpha}(t)\frac{1-\alpha}{\alpha+1}\left(\gamma_{\alpha}(t)\right)^{\frac{1+\alpha}{1-\alpha}}\right]_0^1-\frac{1-\alpha}{\alpha+1}\int_0^1\dot{\gamma}_{\alpha}(t)\left(\gamma_{\alpha}(t)\right)^{\frac{1+\alpha}{1-\alpha}}\,\mathrm{d} t \right)\\
&=& \left(\frac{1-\alpha}{2}\right)^{\frac{2\alpha}{1-\alpha}}\Tr\left(\left[t\dot{\gamma}_{\alpha}(t)\frac{1-\alpha}{1+\alpha}\left(\gamma_{\alpha}(t)\right)^{\frac{1+\alpha}{1-\alpha}}\right]_0^1-\frac{1-\alpha}{1+\alpha}\frac{1-\alpha}{2}\left[\left(\gamma_{\alpha}(t)\right)^{\frac{2}{1-\alpha}}\right]_0^1 \right)\,.
\end{eqnarray*}
At this point we can use Eq. (\ref{alpha-geod}) to obtain
\begin{eqnarray*}
\Div(\rho_1,\rho_2)&=&\left(\frac{1-\alpha}{2}\right)^{\frac{2\alpha}{1-\alpha}} \Tr\Bigg(\frac{2}{1-\alpha}\left(\rho_2^{\frac{1-\alpha}{2}}-\rho_1^{\frac{1-\alpha}{2}}\right)\frac{1-\alpha}{1+\alpha}\left(\frac{2}{1-\alpha}\right)^{\frac{1+\alpha}{1-\alpha}}\rho_2^{\frac{1-\alpha}{2}\frac{1+\alpha}{1-\alpha}}\\
&&-\frac{1-\alpha}{1+\alpha}\frac{1-\alpha}{2}\left(\frac{2}{1-\alpha}\right)^{\frac{2}{1-\alpha}}\left(\rho_2^{\frac{1-\alpha}{2}\frac{2}{1-\alpha}}-\rho_1^{\frac{1-\alpha}{2}\frac{2}{1-\alpha}}\right) \Bigg)\\
&=& \frac{4}{1-\alpha^2}\Tr\left(\left(\rho_2^{\frac{1-\alpha}{2}}-\rho_1^{\frac{1-\alpha}{2}}\right)\rho_2^{\frac{1+\alpha}{2}}\right)- \frac{1-\alpha}{1+\alpha}\frac{2}{1-\alpha}\Tr\left(\rho_2-\rho_1\right)\\
&=& \frac{4}{1-\alpha^2}\Tr\rho_2-\frac{2}{1+\alpha}\Tr\rho_2+\frac{2}{1+\alpha}\Tr\rho_1-\frac{4}{1-\alpha^2}\Tr\left(\rho_1^{\frac{1-\alpha}{2}}\rho_2^{\frac{1+\alpha}{2}}\right)\\
&=& \frac{2}{1+\alpha}\Tr\rho_1+\frac{2}{1-\alpha}\Tr\rho_2-\frac{4}{1-\alpha^2}\Tr\left(\rho_1^{\frac{1-\alpha}{2}}\rho_2^{\frac{1+\alpha}{2}}\right)\,.
\end{eqnarray*}

Finally, we can conclude that
\begin{equation}
\label{alpha-divergence}
\Div(\rho_1,\rho_2)=\frac{4}{1-\alpha^2}\Tr\left(\frac{1-\alpha}{2}\,\rho_1+\frac{1+\alpha}{2}\,\rho_2-\rho_1^{\frac{1-\alpha}{2}}\rho_2^{\frac{1+\alpha}{2}}\right)\,,\quad \rho_1,\,\rho_2\,\in\mathrm{M}_+\,,
\end{equation}
which corresponds to the $\alpha$-canonical divergence on the manifold of  positive definite matrices.

\section{Conclusion}

The present article is a follow-up of the paper \cite{Felice19} recently published. In the latter one, the authors showed that the canonical divergence defined in Eq. \eqref{AyDivnice} provides a powerful tool for unifying classical and quantum Information Geometry. In particular, such a divergence has been proved to coincide with the Kullback-Leibler divergence on the simplex of probability distributions and with the quantum relative entropy on the manifold of quantum states. The effectiveness of the Kullback-Leibler divergence is ascribed in the context of complex systems for quantifying how much a probability measure deviates from the non-interacting states that are modeled by exponential families of probabilities \cite{Aycomplexity}. On the other hand, the quantum relative entropy turns out to be relevant for providing a measure of the many-party
correlations of a quantum state from a Gibbs family \cite{Ayquantum} which in turn is related to the entanglement of quantum systems as defined in \cite{Vedral}.

The $\alpha$-divergence \eqref{alphadivergenceC} is a generalization of the relative entropy \eqref{KL}. Moreover, the flat $\alpha$-geometry induced by the $\alpha$-divergence on the manifold of positive measures constitutes a generalization of the dually flat structure given by the Fisher metric and the mixture and exponential connections. Very remarkably, the $\alpha$-geometry covers the geometry of $\mathrm{q}$-entropy physics \cite{Ohara}. This bridges a very nice connection between the $\alpha$-geometry and the generalized statistical mechanics established by Tsallis \cite{Tsallis,Tsallis19}. On the quantum side, the $\alpha$-divergence \eqref{QalphaDiv} has been introduced as the generalization of the $\alpha$-divergence \eqref{alphadivergenceC} of the positive measures \cite{Amari10}.  The quantum $\alpha$-geometry of the manifold of positive definite matrices is flat, as well, and turns out to be a generalization of the quantum geometry given by the quantum Fisher metric and the quantum mixture and exponential connections induced on the manifold of quantum states by the BKM inner product \cite{Grasselli04}. Also in the quantum setting, the connection between the $\alpha$-geometry and the $\mathrm{q}$-entropy physics by Tsallis bridges a physical interpretation of the $\alpha$-divergence. Indeed, a conditional form of the $\mathrm{q}$-entropy has been employed for the exact calculation, on some systems, of the separable-entangled separatrix \cite{Tsallis}.

In this article, we computed the canonical divergence \eqref{AyDivnice} for flat $\alpha$-connections on the manifold of  positive measures as well as on the manifold of positive definite Hermitian operators. Therefore, we proved that the divergence introduced by Ay and Amari in \cite{Ay15} reduces to the classical $\alpha$-divergence \eqref{alphadivergenceC} and to the quantum $\alpha$-divergence \eqref{QalphaDiv}. Actually, the equivalence between the canonical divergence \eqref{AyDivnice} and the classical $\alpha$-divergence was primarily showed in \cite{Ay15}. {There, the derivation of $D^{(\alpha)}$ was grounded on the idea of a squared distance function associated to the $\alpha$-connections through the vector field $\nihat_t(p)$ given in Eq. \eqref{AyDiv}. However, the $\alpha$-divergence $D^{(\alpha)}$ does not share all the properties of a squared distance function, unless $\alpha=0$. For this reason, in the present paper we obtained the equivalence between $\Div$ and $D^{(\alpha)}$ in a different way, by considering the $\alpha$-divergence as a function of a more general structure.} 

The present paper is conceived within a project which aims to characterize a general canonical divergence for a given dualistic structure $(\metric,\nabla,\nabla^*)$ of a smooth manifold $\Ma$. This project started with the work by Ay and Amari \cite{Ay15} and later developed in \cite{Ay17} where the concept of the ``canonical divergence" has been clearly depicted.  A considerable effort towards the definition of a general canonical divergence has been put forward in \cite{FeliceCanDiv} where a divergence function has been introduced through an extensive investigation of the geodesic geometry of the dualistic structure $(\metric,\nabla,\nabla^*)$.  Actually, this latter divergence turns out to coincide with the divergence \eqref{AyDivnice} when the dualistic structure $(\metric,\nabla,\nabla^*)$ is flat. Further work around this topic is presented in \cite{FeliceGsi} where the very recent divergence is compared together with other divergence functions present in the literature. {It is commonly accepted that the $\alpha$-divergence on the simplex  is obtained by restricting \eqref{alphadivergenceC} to the set of normalized positive measures (see \cite{Amari00} and \cite{Amari16} for more details). Then, the $\alpha$-divergence of  probability distributions in the simplex reads as follows,
\begin{equation}
\label{alphadivesimplex}
D^{(\alpha)}(p,q)=\sum_{i=1}^n\left(1-\frac{4}{1-\alpha^2}p_i^{\frac{1-\alpha}{2}}\,q_i^{\frac{1+\alpha}{2}}\right)\,.
\end{equation}
One can easily verify that for $\alpha=0$, this divergence is closely related to the Hellinger distance which is the distance in the ambient space $\mathcal{M}_+$ \cite{Ay17}.
However, when $\alpha=0$, the $\alpha$-connection is the Levi-Civita connection of the Fisher metric. Therefore, in this case, the canonical divergence has to be one half the squared of the Fisher distance which is different from $D^{(0)}$ \cite{Ay17}, as discussed in Section \ref{CanDic}.   It would be interesting to evaluate the very recent divergence introduced in \cite{FeliceCanDiv} on the simplex of probability distributions, which is not $\alpha$-flat, and to compare it with the $\alpha$-divergence \eqref{alphadivesimplex}.

On the quantum side,} an intriguing work within the above mentioned project is to consider that very recent divergence function, introduced in \cite{FeliceCanDiv}, on the manifold of pure quantum states, where a dually flat structure does not exist \cite{AyTusch}. This will constitute the object of study of a forthcoming investigation.

%%%%%%%%%%%%%%%%%%%%%%%%%%%%%%%%%%%%%%%%%%
%\section{Patents}
%This section is not mandatory, but may be added if there are patents resulting from the work reported in this manuscript.

%%%%%%%%%%%%%%%%%%%%%%%%%%%%%%%%%%%%%%%%%%
\vspace{6pt} 

%%%%%%%%%%%%%%%%%%%%%%%%%%%%%%%%%%%%%%%%%%
%% optional
%\supplementary{The following are available online at \linksupplementary{s1}, Figure S1: title, Table S1: title, Video S1: title.}


%%%%%%%%%%%%%%%%%%%%%%%%%%%%%%%%%%%%%%%%%%
%\authorcontributions{The authors have equally contributed to the manuscript. They all have read and approved
its final version.}

%%%%%%%%%%%%%%%%%%%%%%%%%%%%%%%%%%%%%%%%%%
%\funding{This research received no external funding.}

%%%%%%%%%%%%%%%%%%%%%%%%%%%%%%%%%%%%%%%%%%
\section*{Acknowledgments} 
The authors are very grateful to S.-I. Amari for useful discussions.

%%%%%%%%%%%%%%%%%%%%%%%%%%%%%%%%%%%%%%%%%%
%\conflictsofinterest{The authors declare no conflict of interest.} 

\vspace{6pt} 
%%%%%%%%%%%%%%%%%%%%%%%%%%%%%%%%%%%%%%%%%%
%% optional
%\abbreviations{The following abbreviations are used in this manuscript:\\

%\noindent 
%\begin{tabular}{@{}ll}
%MDPI & Multidisciplinary Digital Publishing Institute\\
%DOAJ & Directory of open access journals\\
%TLA & Three letter acronym\\
%LD & linear dichroism
%\end{tabular}}

%%%%%%%%%%%%%%%%%%%%%%%%%%%%%%%%%%%%%%%%%%
%% optional
%\appendixtitles{no} %Leave argument "no" if all appendix headings stay EMPTY (then no dot is printed after "Appendix A"). If the appendix sections contain a heading then change the argument to "yes".
%\appendix
%\section{}
%\unskip
%\subsection{}
%The appendix is an optional section that can contain details and data supplemental to the main text. For example, explanations of experimental details that would disrupt the flow of the main text, but nonetheless remain crucial to understanding and reproducing the research shown; figures of replicates for experiments of which representative data is shown in the main text can be added here if brief, or as Supplementary data. Mathematical proofs of results not central to the paper can be added as an appendix.

%\section{}
%All appendix sections must be cited in the main text. In the appendixes, Figures, Tables, etc. should be labeled starting with `A', e.g., Figure A1, Figure A2, etc. 


%\section*{References}


\begin{thebibliography}{999}

\bibitem{Amari00}
Amari, S.-I., Nagaoka, H.: Methods of Information Geometry. Oxford University Press (2000)

\bibitem{Felice18}
Felice., D., Cafaro, C., Mancini, S., {\it Information geometric methods for complexity}. Chaos: An Interdisciplinary Journal of Nonlinear Science {\bf 28} (3), 032101 (2018)


\bibitem{Ay17}
Ay, N., Jost, J., Van Le, H., Schwachh\"ofer, L.: {Information Geometry,} 1st ed.; Springer International Publishing: Cham, Switzerland, 2017.

\bibitem{Lauritzen87}
Lauritzen, S.L., {\it Differential Geometry in Statistical Inference},
{Lecture Notes-Monograph Series} {\bf 10},
{163-218} (1987)

\bibitem{Amari82}
Amari, S.-I., {\it Differential geometry of curved exponential families-curvatures and information loss}, Ann. Statist. {\bf 10}, 357-387 (1982).

\bibitem{Amari16}
Amari, S.-I.: Information Geometry and its applications, Springer Japan (2016)

\bibitem{Eguchi85}
Eguchi, S., {\it A differential geometric approach to statistical inference on the basis of contrast
functions}, {Hiroshima Math. J.} {\bf 15},  341--391 (1985)

\bibitem{Eguchi92}
Eguchi, S. {\it  Geometry of minimum contrast}, {Hiroshima Math. J.} {\bf 22}, 631--647 (1992).

\bibitem{Fujiwara95}
Fujiwara, A., Amari, S.-i., {\it Gradient systems in view of information geometry}, Physica D {\bf 80}, 317-327 (1995)

\bibitem{Nakamura93}
Nakamura, Y., {\it Completely integrable gradient systems on the
manifolds of Gaussian and multinomial distributions}, Jap. J.
Industrial and Appl. Math. {\bf 10}, 179 (1993) 

\bibitem{Aycomplexity}
Ay, N., {\it Information geometry on complexity and stochastic interaction}, Entropy {\bf 17}, 2432-2458 (2015)

\bibitem{Nagaoka95}
Nagaoka, H. Differential Geometrical Aspects of Quantum State Estimation and Relative Entropy. In {\it Quantum Communications and Measurement}; Belavkin V.P., Hirota O., Hudson R.L. Eds.; Springer: Boston, MA, USA, 1995.

\bibitem{Ayquantum}
Weis, S., Knauf, A., Ay, N., Zhao, M.J., {\it Maximizing the divergence from a hierarchical model of quantum
states}, Open Syst. Inf. Dyn. {\bf 22}, 1550006 (2015)

\bibitem{Niekamp}
Niekamp, S., Galla, T., Kleinmann, M., G\"uhne, O., {\it Computing complexity measures for quantum states based on exponential families}, J. Phys. A Math. Theor. {\bf 46}, 125301 (2013)

\bibitem{Vedral}
Vedral, V., Plenio, M.B., Rippin, M.A., Knight, P.L., {\it Quantifying entanglement}, Phys. Rev. Lett. {\bf 78}, 2275-2279 (1997)

\bibitem{Grasselli04}
Grasselli, M. H.,
{\it Duality, monotonicity and the Wigner-Yanase-Dyson metrics}. Infinite Dimensional Analysis, Quantum Probability and Related Topics, {\bf 7} (2), 215-232, 2004

\bibitem{Tsallis}
Tsallis, C.: Introduction to nonextensive Statistical Mechanics, Springer (2009)

\bibitem{Tsallis19}
Tsallis, C., {\it Beyond Boltzmann-Gibbs-Shannon in Physics and Elsewhere}, Entropy  {\bf 21}, 696 (2019)

\bibitem{Ohara}
Ohara, A., {\it Geometry of distributions associated with Tsallis statistics and properties of relative entropy minimization}, Physics Letters A {\bf 370}, 184-193 (2007)

\bibitem{Abe}
Abe, S., Rajagopal, A. K., {\it Nonadditive conditional entropy and its significance for local realism}, Physica A {\bf 249}, 157 (2001)

\bibitem{Ay15}
Ay, N., Amari, S.-I., {\it A Novel Approach to Canonical Divergences within
Information Geometry}. { Entropy} {\bf 7},  8111--8129 (2015)

\bibitem{matumoto1993}
Matumoto, T., {\it Any statistical manifold has a contrast function---on the $C\sp 3$-functions taking the minimum at the diagonal of the product manifold}, { Hiroshima Math. J.}  {\bf 23}, 327--337 (1993).


\bibitem{AyTusch}
Ay, N.; Tuschmann, W., {\it Duality versus dual flatness in quantum information geometry}, { J. Math. Phys.}  {\bf 44}, 1512--1518 (2003).



\bibitem{Felice19}
Felice, D., Mancini, S., Ay, N., {\it Canonical divergence for measuring classical and quantum complexity}, Entropy {\bf 21}, 435 (2019)



\bibitem{Centsov}
\v{C}encov, N. N., {\it Statistical decision rules and optimal inference}, American Mathematical Society, Providence, R.I., 1982. Translation from the
Russian edited by Lev J. Leifman.



\bibitem{Petz&Sudar}
Petz, D., Sud\`ar, C., {\it Geometries of quantum states}, Journal of Mathematical Physics {\bf 37}, 2662 (1996)

\bibitem{Holevo}
Holevo, A.S., {\it Probabilistic and Statistical Aspects of Quantum Theory}, North-Holland, Amsterdam, 1982.

\bibitem{Petz96}
Petz, D., {\it Monotone metrics on matrix spaces}, Linear Algebra Appl. {\bf 244}, 81-96 (1996)



\bibitem{Grasselli01}
Grasselli, M. R. ,   Streater, R. F., {\it On the uniqueness of the Chentsov
metric in quantum information geometry}, Inn. Dimens. Anal. Quantum Probab. Relat. Top. {\bf 4} (2), 173-182 (2001) 


\bibitem{Hasegawa97}
Hasegawa, H., {\it Exponential and mixture families in quantum statistics:
dual structure and unbiased parameter estimation}, Rep. Math. Phys.
{\bf 39}(1), 49-68 (1997).

\bibitem{Jencova01}
Jen\v{c}ov\`a, A., {\it Geometry of quantum states: dual connections and divergence
functions}, Rep. Math. Phys. {\bf 47} (1), 121-138 (2001).

\bibitem{Hasegawa92}
Hasegawa, H., {\it $\alpha$-divergence of the noncommutative information geometry},
In Proceedings of the XXV Symposium on Mathematical Physics
(Tor\`un, 1992), volume 33, pages 87-93 (1993).





\bibitem{Martins09}
Martins, A. F. T., Smith, N. A., Xing, E. P., Aguiar, P. M. Q., Figueiredo, M. A. T., {\it Nonextensive Information Theoretic Kernels on Measures}, Journal of Machine Learning Research {\bf 10}, 935-975 (2009)

\bibitem{Jencova03}
Jen\v{c}ov\`a, A., {\it Flat connections and Wigner-Yanase-Dyson metrics}, 	
Reports on Mathematical Physics {\bf 52} (3), 331-351 (2003)

\bibitem{Jancova02}
Jen\v{c}ov\`a, A., {\it Quantum information geometry and standard purification}, J. Math. Phys. {\bf 43} (5), 2187 (2002)

\bibitem{Amari10}
Amari, S.-I., Cichocki, A., {\it Information geometry of divergence functions}, Bulletin of the Polish Academy of Sciences: Technical Sciences {\bf 58} (1), 183-195 (2010)


\bibitem{Furuichi04}
{Furuichi,S., Yanagi,K.,  Kuriyama,K. },
{\it Fundamental properties of Tsallis relative entropy},
{Journal of Mathematical Physics} {\bf 45} (12),
{4868-4877} (2004)

\bibitem{FeliceCanDiv}
Felice, D., Ay, N.: Towards a canonical divergence within Information Geometry. arXiv:1806.11363 [math.DG] (2018)

\bibitem{FeliceGsi}
Felice, D., Ay, N.: Divergence functions in Information Geometry. arXiv:1903.02379 [math.DG] (2019)

\end{thebibliography}
\end{document}